\documentclass[twocolumn,twoside,slac_two]{revtex4}
\usepackage{graphicx}
\usepackage{fancyhdr}
\begin{document}

\title{VERITAS Observations of Blazars}


\author{W. Benbow for the VERITAS Collaboration}
\affiliation{Harvard-Smithsonian Center for Astrophysics, F.L. Whipple Observatory, PO Box 6369, Amado, AZ 85645, USA}

\begin{abstract}

The VERITAS array of four 12-m diameter imaging atmospheric-Cherenkov telescopes in southern Arizona is
used to study very high energy (VHE; E$>$100 GeV) $\gamma$-ray emission from astrophysical objects. 
VERITAS is currently the most sensitive VHE  $\gamma$-ray observatory in the world and one of
the VERITAS collaboration's Key Science Projects (KSP) is the study of blazars. These active galactic
nuclei (AGN) are the most numerous class of identified VHE sources, with $\sim$30 known to emit VHE
photons. More than 70 AGN, almost all of which are blazars, have been observed with the VERITAS
array since 2007, in most cases with the deepest-ever VHE exposure. These observations have
resulted in the detection of VHE $\gamma$-rays from 16 AGN (15 blazars), 
including 8 for the first time at these energies. 
The VERITAS blazar KSP is summarized in this proceeding 
and selected results are presented.

\end{abstract}

\maketitle

\thispagestyle{fancy}


\section{Introduction}
Active galactic nuclei are the most numerous class of identified VHE $\gamma$-ray sources.  
These objects emit non-thermal radiation across $\sim$20 orders of magnitude in energy and
rank among the most powerful particle accelerators in the universe.
A small fraction of AGN possess strong collimated outflows (jets)
powered by accretion onto a supermassive black hole (SMBH).  VHE $\gamma$-ray emission 
can be generated in these jets,  likely in a compact region very near the SMBH event 
horizon.  Blazars, a class of AGN with jets pointed along 
the line-of-sight to the observer, are of particular interest in the VHE regime.
Approximately 30 blazars, primarily high-frequency-peaked BL\,Lacs (HBL), 
are identified as sources of VHE $\gamma$-rays,
and some are spectacularly variable on time scales  
comparable to the light crossing time of their SMBH ($\sim$2 min; \citep{HESS_2155}).  
VHE blazar studies probe the environment very near the central SMBH and
address a wide range of physical phenomena, including the accretion and jet-formation processes.
These studies also have cosmological implications, as VHE blazar data can be used
to strongly constrain primordial radiation fields (see the extragalactic background
light (EBL) constraints from, e.g., \citep{HESS_EBL1, HESS_EBL2}).

VHE blazars have double-humped spectral energy distributions (SEDs), 
with one peak at UV/X-ray energies
and another at GeV/TeV energies.  The origin of the lower-energy peak
is commonly explained as synchrotron emission from the relativistic electrons 
in the blazar jets.  The origin of the higher-energy peak is
controversial, but is widely believed to be the result of inverse-Compton scattering
of seed photons off the same relativistic electrons.  The origin of the seed photons
in these leptonic scenarios could be the synchrotron photons themselves, or photons
from an external source.  Hadronic scenarios
are also plausible explanations for the VHE emission, but generally are not favored.  

Contemporaneous multi-wavelength (MWL) observations of VHE blazars, 
can measure both SED peaks and are crucial 
for extracting information from the observations of VHE blazars.  They are used to 
constrain the size, magnetic field and Doppler factor
of the emission region, as well as to determine the origin (leptonic or hadronic) 
of the VHE $\gamma$-rays.  In leptonic scenarios, such MWL observations are
used to measure the spectrum of high-energy electrons producing 
the emission, as well as to elucidate the nature of the seed photons.  
Additionally, an accurate measure of the 
cosmological EBL density requires accurate modeling of the blazar's 
intrinsic VHE emission that can only be performed
with contemporaneous MWL observations. 

\section{VERITAS}

VERITAS, a stereoscopic array of four 12-m atmospheric-Cherenkov
telescopes located in Arizona, is used to study VHE $\gamma$-rays 
from a variety of astrophysical sources \citep{Holder_AIP}.  VERITAS began scientific observations 
with a partial array in September 2006 and has routinely observed with the 
full array since September 2007. The performance metrics of VERITAS include 
an energy threshold of $\sim$100 GeV, an energy resolution of $\sim$15\%, 
an angular resolution of $\sim$0.1$^{\circ}$, and a sensitivity yielding
a 5$\sigma$ detection of a 1\% Crab Nebula flux object in $<$30 
hours\footnote{A VERITAS telescope was relocated
during Summer 2009, increasing the array's sensitivity 
by a factor $\sim$1.3.}.  VERITAS has an active maintenance program (e.g.
frequent mirror re-coating and alignment) to ensure its continued high performance
over time, and an upgrade improving both the camera (higher quantum-efficiency PMTs) 
and the trigger system has been proposed to the funding agencies.

\section{VERITAS Blazar KSP}

VERITAS observes for $\sim$750 h and $\sim$250 h each year during 
periods of astronomical darkness and partial moonlight, respectively. 
The moonlight observations are almost exclusively used for a blazar
discovery program, and a large fraction of the dark time is used for the 
blazar KSP, which consists of:

\begin{itemize}

\item{A VHE blazar discovery program ($\sim$200 h / yr): Each year $\sim$10 targets are selected to
receive $\sim$10 h of observations each during astronomical darkness. These data are 
supplemented by discovery observations during periods of partial moonlight.}

\item{A target-of-opportunity (ToO) observation program ($\sim$50 h / yr): 
VERITAS blazar observations can be triggered by either a VERITAS blazar discovery, 
a VHE flaring alert ($>$2 Crab) from the blazar monitoring program of the Whipple 10-m 
telescope or from another VHE instrument, or a lower-energy flaring alert
(optical, X-ray or Fermi-LAT). Should the guaranteed allocation 
be exhausted, further time can be requested from a pool of director's discretionary time.}

\item{Multi-wavelength (MWL) studies of VHE blazars ($\sim$50 h / yr + ToO): Each year one
blazar receives a deep exposure in a pre-planned campaign of extensive, simultaneous 
MWL (X-ray, optical, radio) measurements. ToO observation proposals for MWL 
measurements are also submitted to lower-energy observatories (e.g. Swift)
and are triggered by a VERITAS discovery or flaring alert.}

\item{Distant VHE blazar studies to constrain the extragalactic background light (EBL):
Here distant targets are given a higher priority in the blazar discovery program, 
as well as for the MWL observations of known VHE blazars, particularly those 
with hard VHE spectra.}

\end{itemize}

\section{Blazar Discovery Program}

The blazars observed in the discovery program are largely high-frequency-peaked BL Lac objects. 
However, the program also includes IBLs (intermediate-peaked) and LBLs (low-peaked), as
well as flat spectrum radio quasars (FSRQs), in an attempt to increase the
types of blazars known to emit VHE $\gamma$-rays. The observed targets are
 drawn from a {\it target list} containing objects visible to the
telescopes at reasonable zenith angles ($-8^{\circ} < \delta < 72^{\circ}$),
without a previously published VHE limit below
1.5\% Crab, and with a measured redshift $z < 0.3$. To further the study of the
EBL a few objects having a large ($z > 0.3$) are also included in the target list. 
The target list includes:

\begin{itemize}
\item{All nearby ($z < 0.3$) HBL and IBL recommended as potential VHE emitters in \citep{poster1,poster2, poster3}.}
\item{The X-ray brightest HBL ($z < 0.3$) in the recent Sedentary \citep{poster4} and ROXA \citep{poster5} surveys.}
\item{Four distant ($z > 0.3$) BL Lac objects recommended by \citep{poster1,poster6}.}
\item{Several FSRQ recommended as potential VHE emitters in \citep{poster2,poster9}.}
\item{All nearby ($z < 0.3$) blazars detected by EGRET \citep{poster7}.}
\item{All nearby ($z < 0.3$) blazars contained in the Fermi-LAT Bright AGN Sample \citep{poster8}.}
\item{All sources ($|b| > 10^{\circ}$) detected by Fermi-LAT where
extrapolations of their MeV-GeV $\gamma$-ray spectrum
(including EBL absorption; assuming z = 0.3 if the redshift is unknown)
indicates a possible VERITAS detection in less than 20 h.  This criteria
is the focus of the 2009-10 VERITAS blazar discovery program.}
\end{itemize}

\section{VERITAS AGN Detections}

VERITAS has detected VHE $\gamma$-ray emission from 16 AGN (15 blazars),
including 8 VHE discoveries. These AGN are shown in Table~\ref{Blazar_list},
and each has been detected
by the Large Area Telescope (LAT) instrument aboard the Fermi Gamma-ray Space Telescope.
Every blazar discovered by VERITAS was the subject of ToO MWL observations
to enable modeling of its simultaneously-measured SED.  
The known VHE blazars detected by VERITAS were similarly the targets
of MWL observations.

\begin{table}[t]
\begin{center}
\caption{VERITAS AGN Detections.  The only non-blazar
object is the radio galaxy M\,87.  The blazars discovered at VHE by
VERITAS are marked with a dagger.}
\begin{tabular}{|c|c|c|}
\hline \textbf{Object} & \textbf{Class} & \textbf{Redshift} \\
\hline M\,87 & FR I & 0.004 \\
\hline Mkn\,421 & HBL & 0.030 \\
\hline Mkn\,501 & HBL & 0.034 \\
\hline 1ES\,2344+514 & HBL & 0.044  \\
\hline 1ES\,1959+650 & HBL & 0.047 \\
\hline W\,Comae$^{\dagger}$ & IBL & 0.102 \\
\hline RGB\,J0710+591$^{\dagger}$ & HBL & 0.125 \\
\hline H\,1426+428 & HBL & 0.129 \\
\hline 1ES\,0806+524$^{\dagger}$ & HBL & 0.138 \\
\hline 1ES\,0229+200 & HBL & 0.139 \\
\hline 1ES\,1218+304 & HBL & 0.182  \\
\hline RBS\,0413$^{\dagger}$ & HBL & 0.190  \\
\hline 1ES\,0502+675$^{\dagger}$ & HBL & 0.341 \\
\hline 3C\,66A$^{\dagger}$ & IBL & 0.444?  \\
\hline PKS\,1424+240$^{\dagger}$ & IBL & ?  \\
\hline VER\,J0521+211$^{\dagger}$ & ? & ?  \\
\hline
\end{tabular}
\label{Blazar_list}
\end{center}
\end{table}

\subsection{Recent VERITAS Blazar Discoveries}
Prior to the launch of Fermi VERITAS had discovered VHE emission from
2 blazars. These included the first VHE-detected IBL, W\,Comae \citep{WCom1,WCom2},
and the HBL 1ES\,0806+524 \citep{VER_1ES0806}. VERITAS has discovered 6 VHE 
blazars since the launch of Fermi.  Three of these were initially observed
by VERITAS prior to the release of Fermi-LAT results, due to the X-ray brightness 
of the synchrotron peaks of their SEDs.

VHE emission from 3C\,66A was discovered by VERITAS in September 2008 \citep{3C-A} during 
a flaring episode that was also observed by the Fermi-LAT \citep{3C-B}.  The 
observed flux above 200 GeV was 6\% of the Crab Nebula flux and
the measured VHE spectrum was very soft ($\Gamma_{\rm VHE} \sim 4.1$).
RGB\,J0710+591 was detected ($\sim$5.5$\sigma$;
3\% Crab flux above 300 GeV; $\Gamma_{\rm VHE} \sim 2.7$) 
during VERITAS observations from December 2008 to March 2009.
The initial announcement of the VHE discovery \citep{ATel_0710} led to its discovery
above 1 GeV in the Fermi-LAT data using a special analysis.
RBS\,0413, a relatively distant HBL (z=0.19), was observed for 16 h
good-quality live time in 2008-09\footnote{RBS\,0413 was observed further
by VERITAS in Fall 2009.}.  These data resulted in the 
discovery of VHE gamma-rays ($>$270$\gamma$, $\sim$6$\sigma$) at
a flux ($>$200 GeV) of $\sim$2\% of the Crab Nebula flux. The discovery \citep{RBS0413}
was announced simultaneously with the LAT MeV-GeV detection.
The VHE and other MWL observations, including Fermi-LAT data, for each
of these three sources will be the subject of a joint publication 
involving both the VERITAS and LAT collaborations.

\subsection{Discoveries Motivated by Fermi-LAT}

The successful VHE discovery observations by VERITAS of three 
blazars was motivated primarily by 
results from the first year of LAT data taking. In particular,
the VHE detections of PKS\,1424+240 \citep{VERITAS_1424} and
1ES\,0502+675 \citep{VERITAS_0502} were the result of VERITAS observations
triggered by the inclusion of these objects in the Fermi-LAT
Bright AGN List \citep{poster8}.  The former is only the third IBL known
to emit VHE gamma-rays, and the latter is the most distant BL Lac object ($z=0.341$)
detected in the VHE band.  In addition, VER\,J0521+211, likely associated with 
the radio-loud AGN RGB\,J0521.8+2112, was detected by VERTAS 
in $\sim$4 h of observations in October 2009 \citep{VERJ0521}.  These observations 
were motivated by its identification as a $>$30 GeV $\gamma$-ray source in 
the public Fermi-LAT data. Its VHE flux is 5\% of the Crab Nebula flux,
placing it among the brightest VHE blazars detected in recent years.  
VERITAS later observed even brighter VHE flaring 
from VER\,J0521+211 in November 2009 
\citep{VERJ0521_flare}, leading to deeper VHE observations.

\section{Blazars Upper Limits}
\begin{figure*}[!t]
   \centerline{{\includegraphics[width=65mm]{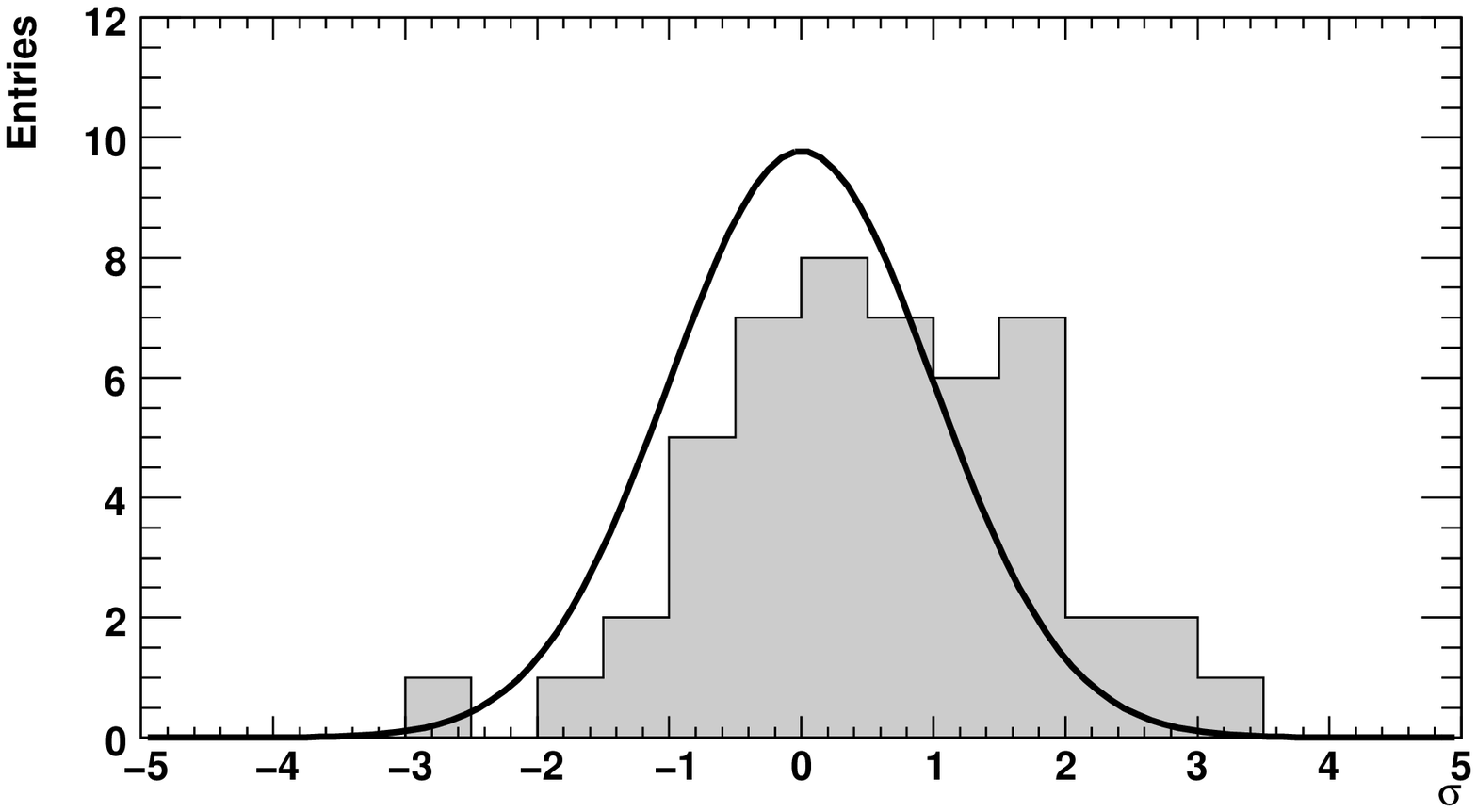}}
              \hfil
              {\includegraphics[width=65mm]{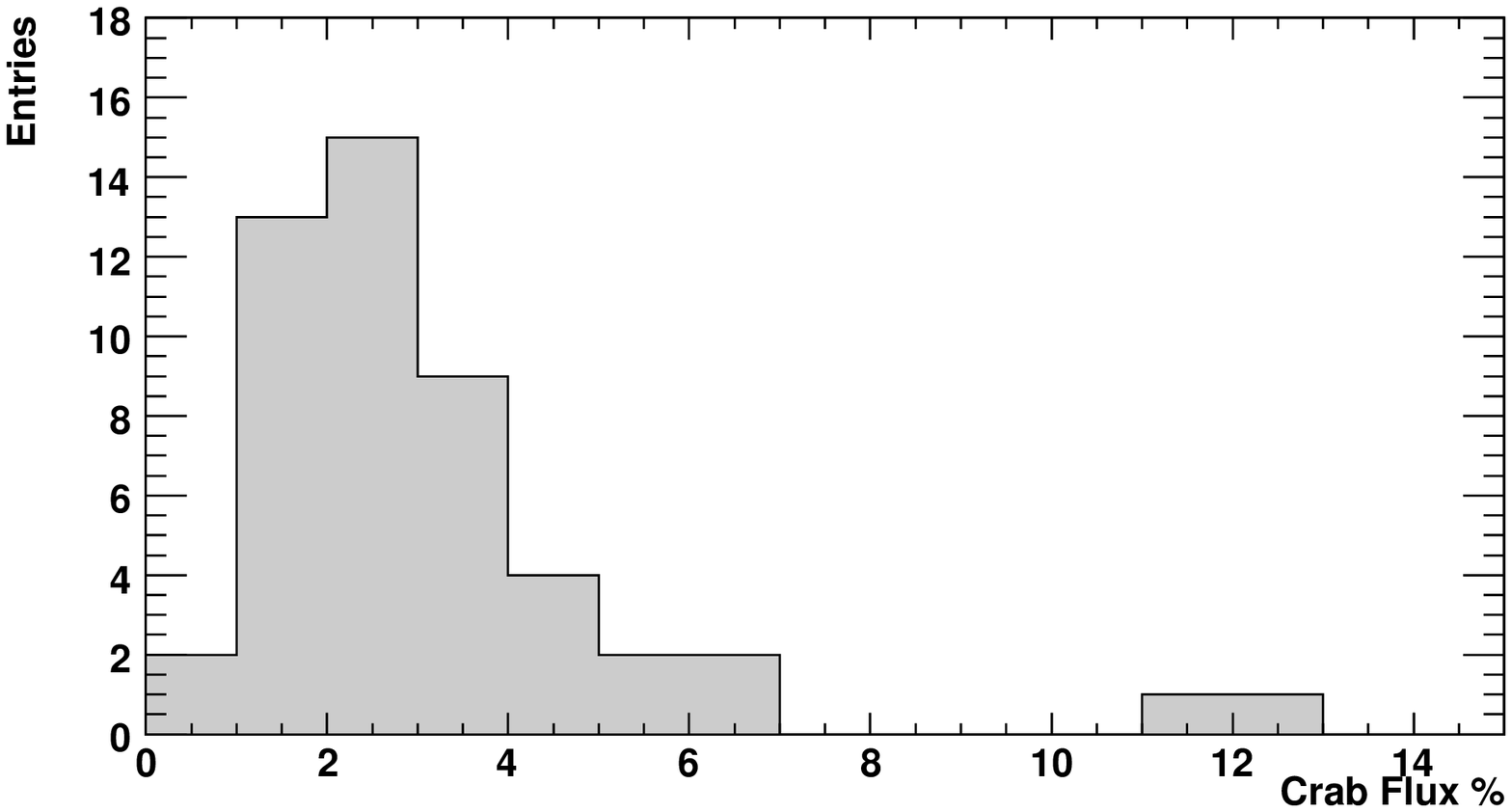}}
             }
   
   \caption{(Left) The preliminary significance measured from each of 
the 49 non-detected candidates using standard analysis cuts. 
The curve shows a Gaussian distribution, with mean zero and 
standard deviation one, normalized to the number of blazars. A similar
result is obtained using analysis cuts optimized for soft-spectrum sources. (Right) 
The distribution of flux upper limits for the non-detected blazars
in percentage of Crab Nebula flux above the observation threshold.  
The time-weighted average
limit is less than $\sim$2\% Crab flux.}
   \label{blazar_UL_plots}
 \end{figure*}

More than 50 VHE blazar candidates were observed by VERITAS
between September 2007 and June 2009. The total exposure 
on the 49 non-detected candidates is $\sim$305 h live time 
(average of 6.2 h per candidate). Approximately 55\% of the 
total exposure is split amongst the 27 observed HBL. The
remainder is divided amongst the 8 IBL (26\%), 5 LBL (6\%),
and 9 FSRQ (13\%). There are no clear indications of 
significant VHE $\gamma$-ray emission from any of
these 49 blazars \citep{Benbow_ICRC09}. However, the observed significance
distribution is clearly skewed towards positive values (see
Figure~\ref{blazar_UL_plots}). A stacking
analysis performed on the entire data sample shows an overall
excess of 430 $\gamma$-rays, corresponding to a statistical significance of
4.8$\sigma$, observed from the directions of the candidate blazars. The
IBL and HBL targets make up 96\% of the observed excess.
Observations of these objects also comprise $\sim$80\% of the total
exposure. An identical stacked analysis of all the extragalactic
non-blazar targets observed, but not clearly detected ($>$5$\sigma$), by
VERITAS does not show a significant excess ($\sim$120 h exposure).
The stacked excess persists using alternate methods for estimating
the background at each blazar location, and with different event
selection criteria (e.g. {\it soft cuts} optimized for sources 
with $\Gamma_{\rm VHE} > 4$).  The distribution of VHE
flux upper limits is shown in Figure~\ref{blazar_UL_plots}.
These 49 VHE flux upper limits are generally the most-constraining
ever reported for these objects.

\section{Multi-wavelength Studies of VHE Blazars}
During the first three seasons of VERITAS observations,
pre-planned extensive MWL campaigns were organized for three
blazars 1ES 2344+514 (2007-08), 1ES 1218+304 (2008-09) 
and 1ES 0229+200 (2009-10 - ongoing).  In addition, 
numerous ToO MWL-observation campaigns were performed.
These include campaigns for every blazar/AGN discovered by
VERITAS, and all include Swift (XRT and UVOT) data.
All MWL campaigns on the VHE blazars discovered
since the launch of Fermi include LAT detections. 
In addition, several MWL campaigns on the well-studied 
VHE blazars Mkn 421 and Mkn 501 (please see the 
contributions of D. Gall and A. Konopelko in these proceedings) 
were also performed. Highlights of these campaigns include:

\begin{itemize}
\item{1ES 2344+514: A major (50\% Crab) VHE flare, along with
correlations of the VHE and X-ray flux were observed from this HBL. The VHE 
and X-ray spectra harden during bright states, and a 
synchrotron self-Compton (SSC) model can explain the observed SED 
in both the high and low states \citep{VER_1ES2344}.}

\item{1ES 1218+304: This HBL flared during VERITAS MWL
observations. Its unusually hard VHE
spectrum strongly constrains the EBL. The observed
flaring rules out kpc-scale jet emission as the
explanation of the spectral hardness and places the
EBL constraints on more solid-footing \citep{VER_1ES1218A,VER_1ES1218B}.}

\item{1ES 0806+524: The observed SED of this new VHE HBL
can be explained by an SSC model \citep{VER_1ES0806}.}

\item{W Comae: This IBL, the first discovered at VHE,
flared twice in 2008 \citep{WCom1,WCom2}. Modeling of the SED is
improved by including an external-Compton (EC)
component in an SSC interpretation.}

\item{3C 66A: This IBL flared at VHE and MeV-GeV energies
in 2008\citep{3C-A,3C-B}. Similar to W Comae and PKS 1424+240,
modeling of observed SED suggests a strong EC component
in addition to an SSC component.}

\item{Mkn 421: This HBL exhibited major flaring behavior for
several months in 2008. Correlations of the
VHE and X-ray flux were observed, along with spectral
hardening with increased flux in both bands \citep{Mkn421}.}

\item{RGB\,J0710+591: Modeling the SED of this HBL with an
SSC model yields a good fit to the data.  The inclusion of
an external Compton component does not improve the fit.}

\item{PKS\,1424+240: The broadband SED of this IBL (at unknown redshift) is well described 
by an SSC model favoring a redshift of less than 0.1 \citep{VERITAS_1424}. Using the photon 
index measured with Fermi-LAT in combination with recent 
EBL absorption models, the VERITAS data indicate that the redshift
of PKS 1424+240 is less than 0.66.}

\end{itemize}

\section{Conclusions}
The first two years of the VERITAS blazar KSP were highly successful. Highlights include the
detection of more than a 16 VHE blazars with the observations almost
always having contemporaneous MWL data. Among these detections are
8 VHE blazar discoveries, including the first three IBLs known to emit VHE
$\gamma$-rays.  All but a handful of the blazars on the initial VERITAS
discovery {\it target list} were observed, and the flux limits generated for those
not VHE detected are generally the most-constraining ever.
The excess seen in the stacked blazar analysis suggests that the 
initial direction of the VERITAS discovery program was well justified,
and that follow-up observations of many of these initial targets will result
in VHE discoveries.  In addition, the Fermi-LAT is 
identifying many new compelling targets for the VERITAS blazar discovery program.
These new candidates have already resulted in 3 VHE blazar discoveries.
The future of the VERITAS blazar discovery program is clearly
very bright.

The MWL aspect of the VERITAS blazar KSP has also been highly successful.  Every
VERITAS observation of a known, or newly discovered, VHE blazar has
been accompanied by contemporaneous MWL observations.  These data
have resulted in the identification of correlated VHE and X-ray flux
variability, as well as correlated spectral hardening in both the VHE and X-ray bands.
The VHE MWL observations were performed in both ''quiescent'' and flaring
states for some of the observed blazars.  For the observed HBL objects, the SEDs
can be well described by a simple SSC model in both high and low states.
However, an additional external Compton component is necessary to
adequately fit the SEDs of the IBL objects.  

The Fermi-LAT is already having a significant impact on
the blazar KSP.  In future seasons, the VERITAS blazar discovery program 
will focus its discovery program on hard-spectrum blazars detected by 
Fermi-LAT, and will likely have a greater focus on high-risk/high-reward 
objects at larger redshifts ($0.3 < z < 0.7$). In addition, the number of VHE 
blazars studied in pre-planned MWL campaigns 
will increase as data from the Fermi-LAT will be publicly available. 
In particular,  the extensive pre-planned MWL campaigns will focus on objects that
are noteworthy for the impact their data may have on understanding the EBL.
The simultaneous observations of blazars by VERITAS and Fermi-LAT will
completely resolve the higher-energy SED peak, often for the first time,
enabling unprecedented constraints on the underlying blazar phenomena
to be derived.

\bigskip % extra skip inserted
\begin{acknowledgments}
This research is supported by grants from the US Department of Energy, 
the US National Science Foundation, and the Smithsonian Institution, 
by NSERC in Canada, by Science Foundation Ireland, and by STFC in the UK. 
We acknowledge the excellent work of the technical support staff at the 
FLWO and the collaborating institutions in the 
construction and operation of the instrument.

\end{acknowledgments}


\begin{thebibliography}{99}

\bibitem[1]{HESS_2155}  F.~Aharonian et al. 2007, \emph{ApJ}, {\bf 664}, L71
\bibitem[2]{HESS_EBL1}  F.~Aharonian et al. 2006, \emph{Nature}, {\bf 440}, 1018
\bibitem[3]{HESS_EBL2}  F.~Aharonian et al. 2007, \emph{A\&A}, {\bf 475}, L9
\bibitem[4]{Holder_AIP} J. Holder, et al. 2008, \emph{AIPC}, {\bf 1085}, 657
\bibitem[5]{poster1}    L.~Costamante \& G.~Ghisellini 2002, \emph{A\&A}, {\bf 384}, 56
\bibitem[6]{poster2} E.S.~Perlman 2000, \emph{AIPC}, {\bf 515}, 53
\bibitem[7]{poster3} F.W.~Stecker et al. 1996, \emph{ApJ},  {\bf 473}, L75
\bibitem[8]{poster4} P.~Giommi et al. 2005, \emph{A\&A},  {\bf 434}, 385
\bibitem[9]{poster5} S.~Turriziani et al. 2007, \emph{A\&A},  {\bf 472}, 699
\bibitem[10]{poster6} L.~Costamante 2006, arXiv:0612709
\bibitem[11]{poster9} P.~Padovani et al. 2002, \emph{ApJ},  {\bf 581}, 895
\bibitem[12]{poster7} R.~Muhkerjee et al. 2001, \emph{AIPC}, {\bf 558}, 324
\bibitem[13]{poster8} A.A.~Abdo et al. 2009, \emph{ApJ}, {\bf 700}, 597
\bibitem[14]{WCom1} V.A.~Acciari et al. 2008, \emph{ApJ}, {\bf 684}, L73
\bibitem[15]{WCom2} V.A.~Acciari et al. 2009,  \emph{ApJ}, {\bf 707}, 612
\bibitem[16]{VER_1ES0806} V.A.~Acciari et al. 2009,  \emph{ApJ}, {\bf 690}, L126
\bibitem[17]{3C-A} V.A.~Acciari et al. 2009,  \emph{ApJ}, {\bf 693}, L104
\bibitem[18]{3C-B} L.C.~Reyes 2009, arXiv:0907.5175
\bibitem[19]{ATel_0710} R.A.~Ong 2009,  \emph{ATel}, {\bf 1941}
\bibitem[20]{RBS0413} R.A.~Ong et al. 2009,  \emph{ATel}, {\bf 2272}
\bibitem[21]{VERITAS_1424} V.A.~Acciari et al. 2009,  \emph{ApJ}, {\bf 708}, L100
\bibitem[22]{VERITAS_0502} R.A.~Ong et al. 2009,  \emph{ATel}, {\bf 2301}
\bibitem[23]{VERJ0521} R.A.~Ong et al. 2009,  \emph{ATel}, {\bf 2260}
\bibitem[24]{VERJ0521_flare} R.A.~Ong et al. 2009,  \emph{ATel}, {\bf 2309}
\bibitem[25]{Benbow_ICRC09} W.~Benbow 2009, arXiv:0908.1412
\bibitem[26]{VER_1ES2344} V.A.~Acciari et al. 2009,  \emph{ApJ}, {\bf submitted}
\bibitem[27]{VER_1ES1218A} V.A.~Acciari et al. 2009,  \emph{ApJ}, {\bf 695}, 1370
\bibitem[28]{VER_1ES1218B} V.A.~Acciari et al. 2009, \emph{ApJ}, {\bf in press}
\bibitem[29]{Mkn421} J.~Grube 2009, arXiv:0907.4862

\end{thebibliography}
\end{document}